\documentclass[fleqn,usenatbib]{mnras}

\usepackage[T1]{fontenc}
\usepackage{ae,aecompl}

\usepackage{graphicx}	
\usepackage{amsmath}	
\usepackage{amssymb}	
\usepackage{multirow}
\usepackage{pdflscape}

\def\arcsec{\hbox{$^{\prime\prime}$}}
\def\arcmin{\hbox{$^{\prime}$}}

\newcommand{\casa}{{\tt CASA}}
\newcommand{\wsc}{{\tt WSClean}}

\newcommand{\pybdsf}{{\tt pyBDSF}}
\newcommand{\trap}{{\tt TraP}}

\newcommand{\asec}{$^{\prime\prime}$}
\newcommand{\fhs}{\mbox{\ensuremath{.\!\!^{\rm s}}}}

\newcommand{\fb}{MKT\,J170456.2$-$482100}
\newcommand{\gx}{GX\,339$-$4}

\newcommand{\exofull}{EXO\,040830$-$7134.7}
\newcommand{\exo}{EXO\,0408}

\newcommand{\gudel}{G\"udel}

\newcommand*\oline[1]{%
  \vbox{%
    \hrule height 0.5pt
    \kern0.25ex
    \hbox{%
      \kern-0.1em
      \ifmmode#1\else\ensuremath{#1}\fi
      \kern-0.1em
    }
  }
}

\usepackage{xcolor}

\emergencystretch=\maxdimen
\usepackage{newtxtext,newtxmath}

\title[Detection of radio emission from \exofull]{The detection of radio emission from known X-ray flaring star \exofull}

\author[L. N. Driessen et al.]{L. N. Driessen,$^{1}$\thanks{E-mail: Laura@Driessen.net.au (LND)}
D. R. A. Williams,$^{1}$
I. McDonald,$^{1, 2}$
B. W. Stappers,$^{1}$
\newauthor
D. A. H. Buckley,$^{3,4}$
R. P. Fender,$^{5,6}$
P. A. Woudt,$^{4,6}$
\\ \\
$^{1}$Jodrell Bank Centre for Astrophysics, Department of Physics and Astronomy, The University of Manchester, Manchester, M13 9PL, UK\\
$^{2}$School of Physics and Astronomy, Open University, Walton Hall, Milton Keynes, MK7 6AA, UK\\
$^{3}$South African Astronomical Observatory, PO Box 9, Observatory 7935, South Africa\\
$^{4}$Department of Astronomy, University of Cape Town, Private Bag X3, Rondebosch 7701, South Africa\\
$^{5}$Department of Physics, Astrophysics, University of Oxford, Denys Wilkinson Building, Keble Road, Oxford OX1 3RH, UK\\
$^{6}$Inter-University Institute for Data Intensive Astronomy, Department of Astronomy, University of Cape Town, Private Bag X3,\\Rondebosch 7701, South Africa\\
}
\date{Accepted 2021 November 25. Received 2021 November 25; in original form 2021 August 25}

\pubyear{2021}

\begin{document}
\label{firstpage}
\pagerange{\pageref{firstpage}--\pageref{lastpage}}
\maketitle

\begin{abstract}
We report the detection of radio emission from the known X-ray flaring star \exofull\,during MeerKAT observations of the nearby cataclysmic variable VW Hydri.
We have three epochs of MeerKAT observations, where the star is not detected in the first epoch, is detected in the second epoch, and is marginally detected in the third epoch.
We cannot distinguish whether the detection is quiescent emission or {a transient radio burst}.
If we assume the radio detection is quiescent emission the source lies somewhat to the right of the \gudel--Benz relation;
however, if we assume the upper-limit on the radio non-detection is {indicative of} the quiescent emission then the source lies directly on the relation. {Both cases are broadly consistent with the relation.}
We use archival spectral energy distribution data and new SALT high-resolution spectroscopy to confirm that \exofull\,is a chromospherically active M-dwarf with a temperature of $4000\pm200$\,K of spectral type M0V. We use ASAS, ASAS-SN and \emph{TESS} optical photometry to derive an improved rotational period of 5.18$\pm$0.04\,days. This is the first radio detection of the source, and the first MeerKAT detection of an M-dwarf.
\end{abstract}



\section{Introduction}
\label{sec: EXO intro}

Observing stellar radio variability reveals information about particle acceleration and magnetic fields in stellar atmospheres \citep[e.g.][]{2019ApJ...871..214V,2019PASP..131a6001M} and is also important for determining the habitability of planets orbiting active stars \citep[e.g.][]{2017PhDT.........8V}.
It has been hypothesised that the quiescent radio and X-ray emission from active stars is due to many small flares \citep[e.g.][]{2011ASPC..448..455F,2017PhDT.........8V}. 
{Many cool M-dwarfs tend to be active and are known to host low-mass planets \citep[][]{2015ApJ...807...45D,2019AJ....158...75H}. The population of known active stars and stellar binaries include T-Tauri stars, RS Canum Venaticorum (RS CVn) binaries, Algol binaries, and Cataclysmic Variable (CV) binaries.}
Therefore, it is essential to improve our understanding of the radio nature and variability of main-sequence stars, and thereby the impact of stellar flares on orbiting planets.

{An empirical relation that is often used to determine whether a star is a member of the population of incoherent radio stellar sources is the \gudel--Benz relation. This relation describes a correlation between the radio and X-ray luminosities of stellar radio sources and is given by $L_X/L_R \approx 10^{15.5}$ where $L_R$ and $L_X$ are the quiescent radio specific and X-ray luminosities respectively \citep{1993ApJ...405L..63G}. This relation extends across many magnitudes of X-ray and radio luminosities and applies to solar flares, RS CVn binaries, and BY Draconis-type binaries, as well as active M- and K-type stars. A similar relation exists for Algol systems, FK Com stars, and post T Tauri stars \citep{1994A&A...285..621B}. The \gudel--Benz relation suggests that the mechanism that heats the coronae of active stars is the same mechanism that accelerates non-thermal electrons. The model commonly used to explain this is that magnetic reconnection events in the corona accelerates electrons to produce gyrosynchrotron emission. These non-thermal electrons lose energy in the chromosphere, heating the corona and results in thermal X-ray emission \citep{1994A&A...285..621B}. This means that radio observations of gyrosynchrotron emission from active stars probe the conditions in the stellar corona \citep[e.g.][]{1990SoPh..130..265B}, that cannot be probed at other wavelengths. The gyrosynchrotron emission is detected as incoherent radio emission with low circular polarisation fractions and low brightness temperature ($T_{b}\lesssim10^{12}$\,K).
Stars are also observed to emit coherent radio emission, which is caused by a different emission mechanism, either plasma or electron-cyclotron maser (ECM) emission \citep{1985ARA&A..23..169D}. Coherent radio emission has $T_{b}\gtrsim10^{12}$\,K and a high circular polarisation fraction. Coherent radio emission also violates the \gudel-Benz relation, as recently demonstrated by \citet{2021NatAs.tmp..196C} using Low-Frequency Array \citep[$\lesssim200\,\mathrm{MHz}$][]{} detections of coherent radio emission from a population of M-dwarfs. They found that the radio luminosities of their sources did not correlate with the X-ray luminosity, which suggests that the radio emission is not linked to chromospheric activity. This is particularly interesting as low-frequency, coherent emission from stars could be caused by ECM emission, where there is a magnetic connection between the M-dwarf and an orbiting planet similar to the Jupiter-Io system \citep{2020NatAs...4..577V,2021NatAs.tmp..196C}. Auroral activity on the M-dwarf UV Ceti caused by ECM emission has also been detected at 888\,MHz using the Australian Square Kilometer Array Pathfinder \citep[ASKAP\footnote{\href{https://www.atnf.csiro.au/projects/askap/index.html}{https://www.atnf.csiro.au/projects/askap/index.html}};][]{2020ApJ...905...23Z,2021PASA...38....9H}. These results show that we still have more to learn and discover using radio observations of stellar sources.}

\exofull\,{(hereafter \exo)} was first discovered serendipitously by \citet{1989_exo_discovery} during the European Space Agency's X-ray Observatory \citep[\textit{EXOSAT};][]{1988_EXOSAT_Description} observations of the CV system {VW Hydri (VW Hyi)}. VW Hyi was observed 30 times by \textit{EXOSAT} between 1983 and 1985, for a total of 84.1\,hours. In 29 of the 30 observations, \exo\,was observed as a weak, quiescent X-ray source with a luminosity of $1.8\times10^{29}\,\mathrm{\,erg\,s^{-1}}$ in the soft X-ray band (0.05--2.0\,keV, assuming a distance of 50\,kpc). It was determined by \citet{1989_exo_discovery} that the X-ray source is coincident with a dMe type star of spectral type M0V, because they observe the hydrogen Balmer lines in emission. 
On 1984 October 11, they observed the source to flare with an X-ray flux of $1.2\times10^{30}\,\mathrm{\,erg\,s^{-1}}$. The observation was 7,600\,seconds long, and they did not observe the rise or decay of the flare. At the time, this was the longest X-ray flare observed from a dMe star.
In \citet{1989_exo_discovery} they hypothesise that the X-ray variability is caused by variable active regions on the surface of a rotating star.

{\citet{1996_EXO_firstPeriod} performed} high-precision photometry on active stars discovered with \textit{EXOSAT} using the 50\,cm and 1\,m ESO telescopes at La Silla, Chile. They classified {\exo}\,as an M0Ve star at a distance of 54\,pc. They found some evidence of optical photometric variability with a period of $5.2\pm0.20$\,days.
An updated distance of $57.72\pm0.04$\,pc \citep[][]{2021AJ....161..147B} was provided by the \emph{Gaia} Early Data Release 3 \citep[Gaia EDR3;][]{2020yCat_GAIA_EDR3}.
{The \emph{Gaia} EDR3 J2016.0 position of \exo\,is  04$^{\rm h}$08$^{\rm m}$07\fhs1371$\pm$0\farcs00001 $-$71$^{\circ}$26\arcmin59\farcs1953$\pm$0\farcs00001 in Right Ascension and Declination respectively.}
To the best of our knowledge, \exo\,has not previously been detected in the radio.

ThunderKAT\footnote{The HUNt for Dynamic and Explosive Radio transients with MeerKAT} \citep{ThunderKAT2017} is a MeerKAT Large Survey Project (LSP) observing and investigating variable and transient radio sources such as CVs, X-ray binaries, and $\gamma$-ray bursts.
ThunderKAT also performs commensal searches for variable and transient sources in the field of view (FoV) of MeerKAT observations, taking advantage of the wide ($\gtrsim$1\,square\,degree at 1.4\,GHz) FoV and sensitivity of the telescope.
The (more) Karoo Array Telescope \citep[MeerKAT;][]{2018ApJ...856..180C} is a 64 dish radio telescope in South Africa.
Each dish has an effective area of 13.5\,m and the longest baseline in the array is $\sim$8\,km.

ThunderKAT has agreements to search commensally for variable and transient radio sources in MeerKAT LSP observations. This includes {untargeted} searches, such as the search that led to the discovery of a new transient source, {\fb,} in MeerKAT observations of \gx\,\citep[][]{FlareyBoi}, and looking for possible radio variability from specific sources such as flare stars. To facilitate this, we used catalogues of flare-type objects \citep{2011AJ....142..138L,2013MNRAS.429.2934K,2009ApJ...696..870D,2000A&AS..143....9W,2017ASPC..509....3D} to determine whether they fall into the primary beam of MeerKAT LSP observations\footnote{The code for this can be found here: \href{https://doi.org/10.5281/zenodo.4515114}{https://doi.org/10.5281/zenodo.4515114}}. 
Testing this method on ThunderKAT observations, revealed that \exo\, is within the field of view of MeerKAT in the L-band (856--1712\,MHz) when it was pointed at VW Hyi.

In this paper we will present the discovery of radio emission from \exo\,during the MeerKAT observations of VW Hyi. In Sections\,\ref{sec: EXO MKT obs}, \ref{sec: EXO optical photometry}, \ref{sec: EXO optical spectroscopy}, and \ref{sec: EXO UV} we will respectively present the radio, optical photometric, optical spectroscopic, X-ray, and {Ultraviolet (UV)} observations of \exo. In Sections\,\ref{sec: EXO discussion} and \ref{sec: EXO conclusions} we will discuss and conclude.

\section{MeerKAT radio observations}
\label{sec: EXO MKT obs}

We observed the {CV} VW Hyi on 2018 August 05, 07 and 08 with the MeerKAT radio telescope 
as part of the ThunderKAT
LSP during an outburst. Each observation was two hours, with $\sim$1.5\,hours on target per epoch with a central frequency of 1284\,MHz, a bandwidth of 856\,MHz, 4096 frequency channels and an integration time of 8\,seconds. 61 MeerKAT antennas participated in the first two observations, while 58 antennas were available during the third observation. We interleaved ten-minute scans of the target field with two-minute scans of a bright phase calibrator: PKS\,0408$-$65, which we also used as the flux and band-pass calibrator for all three datasets. 

We reduced the data using the \texttt{Oxkat} \citep{oxkat}\footnote{The analysis scripts can be found at \href{https://github.com/IanHeywood/oxkat}{https://github.com/IanHeywood/oxkat}} MeerKAT reduction scripts, which perform standard calibration and imaging procedures of MeerKAT radio data. These scripts are containerised using \texttt{singularity} \citep{singularity} and \texttt{stimela} \citep{stimela}. We calibrated the data using the 1GC script and \casa\, Version 5.6.2 \citep{CASA} software.  The 1GC script averages the data to 1024 channels, performs standard phase and amplitude calibrations and applies both a priori and automated flagging of the data. We performed additional flagging (FLAG script) before imaging (2GC) of the target field using \wsc\, \citep{wsclean} and \texttt{cubical} \citep{cubical}. We used the default \texttt{Oxkat} settings for MeerKAT data to produce the images. We performed self calibration on the target field and made the final measurement set for each observation.

We made the final images using \wsc\, \citep{offringa-wsclean-2014}, producing full time and frequency integrated images per epoch as well as images with various, shorter integration times from 16\,seconds per image to half an hour per image. {The root-mean-square (RMS) noise of the full integration images was $0.03\,\mathrm{mJy\,beam^{-1}}$ in all three epochs. We also produced eight subband images per epoch with central frequencies 909, 1016, 1123, 1230, 1337, 1444, 1551, and 1658\,MHz.} We measured the Multi-Frequency-Synthesis (MFS) flux density of \exo\,and determined its light curve\footnote{A light curve is a time series of brightness/flux density measurements for a source.} using the LOFAR {\sc Transient Pipeline} \citep[\trap, Release 4.0;][]{Swinbank2015}. \trap\,automatically processes a time series of fits images, detecting sources and extracting light curves. We used the integrated flux density of the source extracted by \trap, where we forced \trap\,to monitor the position of \exo. This means that we have flux density measurements at that location whether \exo\,is detected or not. 

\begin{figure}
\centering
 \includegraphics[width=\columnwidth]{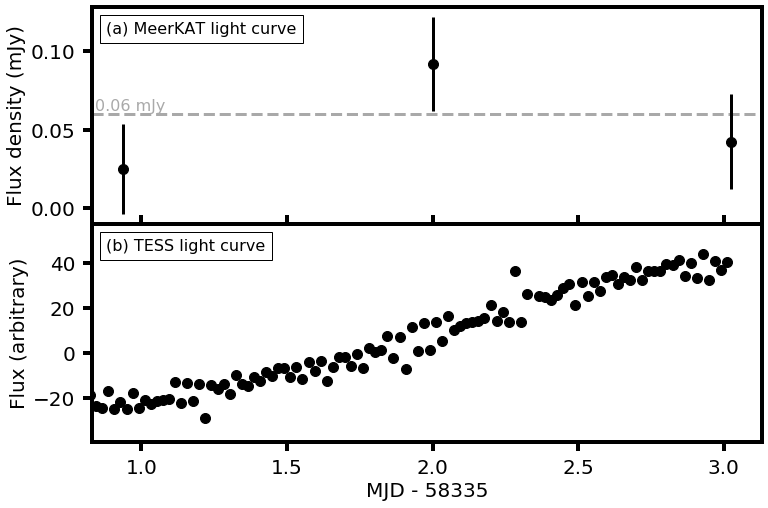}
 \caption[\emph{TESS} and MeerKAT light curves of \exo.]{\emph{TESS} and MeerKAT light curves of \exo. The MeerKAT light curve (a) shows the \trap\,extracted integrated flux density from the full integration images of each epoch. {The grey dashed line indicates three times the image RMS noise level.} The \emph{TESS} light curve (b) has been cleaned and detrended (see Section\,\ref{sec: EXO TESS obs}).}
 \label{fig: EXO MKT TESS lc}
\end{figure}

\begin{figure}
\centering
 \includegraphics[width=\columnwidth]{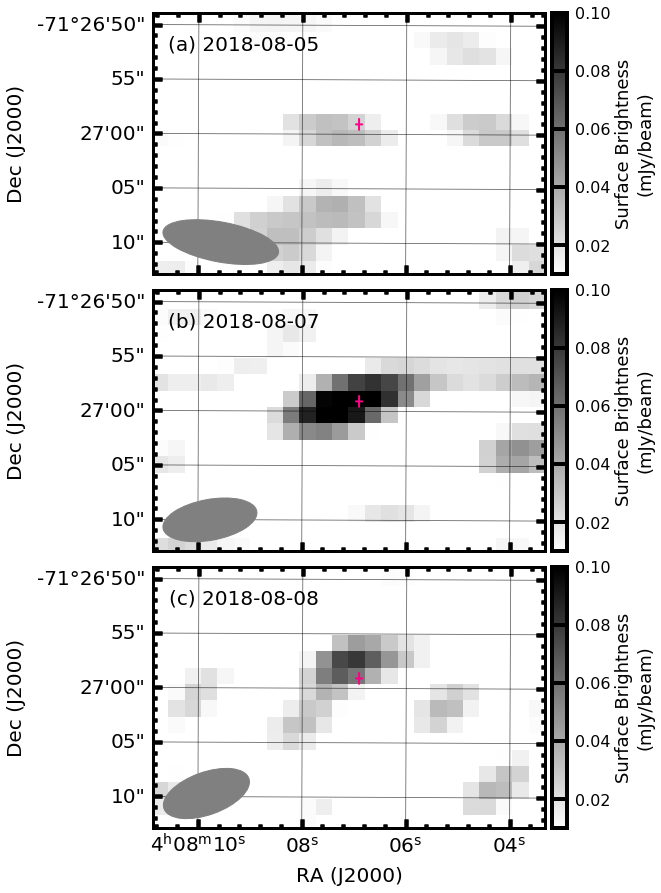}
 \caption{{Full time integration MeerKAT images of \exo. The magenta crosshairs show the \emph{Gaia} position. From top to bottom the epochs are: 2018 August 05 22:26:56.3; 2018 August 07 00:04:13.7; and 2018 August 08 00:37:04.0. The synthesised beam shape is shown in the bottom left corner of each panel, and the greyscale is the same in all three images.}}
 \label{fig: EXO MKT postage stamps}
\end{figure}

The light curve showing the flux density from the full-time-integration images per epoch is shown in Figure\,\ref{fig: EXO MKT TESS lc}(a) and the corresponding images are shown in Figure\,\ref{fig: EXO MKT postage stamps}. The flux density of \exo\,is consistent with zero in the first epoch, the source is detected in the second epoch, and is {marginally detected (S/N$\sim$2) in the third epoch with a position offset of approximately one pixel (1\farcs5). The flux density of the source in each epoch is shown in Table\,\ref{tab: MKT fluxes} as well as the signal-to-noise (S/N) in each epoch.}
Inspection of the shorter-time-integration images revealed no evidence that the source had exhibited a short-duration ($<$1.5\,hr) radio flare and the \trap\,did not detect the source with a signal-to-noise greater than three in any of the shorter integration images.
If we assume the flux density of the source is constant across the three days, the reduced $\chi^{2}$ value is 1.4, which is not sufficient to reject the null hypothesis that the flux density is constant. As such, we cannot confirm whether the source is constant or variable in the radio at this stage. 
{The source is too faint to determine the in-band radio spectral index using the subband images.}

\begin{table}
    \centering
    \begin{tabular}{lrrr}
     & Observation & Flux density ($\mathrm{mJy}$) & S/N \\
    \hline
    \hline
    1 & 2018 August 05 22:26:56.3 & $0.03\pm0.03$ & 1.0 \\
    2 & 2018 August 07 00:04:13.7 & $0.09\pm0.03$ & 4.5 \\
    3 & 2018 August 08 00:37:04.0 & $0.04\pm0.03$ & 2.0 \\
    \hline
    \end{tabular}
    \caption{{The flux density at the position of \exo\,in each MeerKAT full time and frequency integration image. The RMS noise of $0.02\,\mathrm{mJy\,beam^{-1}}$ for each image was used to calculate the S/N.}}
    \label{tab: MKT fluxes}
\end{table}

\subsection{Absolute astrometry}
\label{sec: radio astrometry}

{In approximately the first year of observing, there was a known scatter of at worst 3\arcsec\,on the astrometry of sources detected by MeerKAT. This issue has since been significantly reduced; however, our observations are from early in the first year of MeerKAT observations. As such, we need to perform astrometric corrections to confirm the positions of sources in the field. To do this, we used the Python Blob Detector and Source Finder (\pybdsf\footnote{\href{https://www.astron.nl/citt/pybdsf/}{https://www.astron.nl/citt/pybdsf/}}) to determine the positions of sources in the field in each of the three epochs. We used the positions (see  Table\,\ref{tab: astrometry}) of four Australian Telescope Compact Array (ATCA) Parkes-MIT-NRAO (PMN) \citep[ATPMN;][]{2012MNRAS.422.1527M} sources in the field of view as reference sources. The median astrometric uncertainty of ATPMN sources is 0\farcs4 in both Right Ascension and Declination \citep{2012MNRAS.422.1527M}.}

\begin{table*}
    \centering
    \begin{tabular}{lrrrrr}
    ATPMN source name & ATPMN RA (deg) & ATMPN Dec (deg) & Epoch 1 separation (\asec) & Epoch 2 separation (\asec) & Epoch 3 separation (\asec) \\
    \hline
    \hline
    J040400.5$-$710908 & 04$^{\rm h}$04$^{\rm m}$00\fhs6 & $-$71$^{\circ}$09\arcmin08\farcs9 & 0.6/0.07  & 0.7/0.06  & 0.7/0.07  \\
    J040541.4$-$710449 & 04$^{\rm h}$05$^{\rm m}$41\fhs5 & $-$71$^{\circ}$04\arcmin49\farcs8 & 0.3/0.1   & 0.4/0.1   & 0.3/0.1   \\
    J040748.6$-$705712 & 04$^{\rm h}$07$^{\rm m}$48\fhs7 & $-$70$^{\circ}$57\arcmin12\farcs5 & 0.4/0.05  & 0.4/0.04  & 0.4/0.05  \\
    J040914.4$-$714205 & 04$^{\rm h}$09$^{\rm m}$14\fhs5 & $-$71$^{\circ}$42\arcmin05\farcs3 & 0.8/0.002 & 0.5/0.002 & 0.6/0.002 \\
    \hline
    \end{tabular}
    \caption{{Names and coordinates of the ATPMN reference sources used for absolute astrometry correction and the separation between the ATPMN and MeerKAT reference sources before/after applying the transformation to the source position from each MeerKAT full time and frequency integration image.}}
    \label{tab: astrometry}
\end{table*}

We manually confirmed that the ATPMN sources match unresolved MeerKAT sources and solved for an affine transformation matrix to translate and rotate the MeerKAT source positions and applied this transformation to all of the MeerKAT sources. The separation between the MeerKAT and ATPMN reference sources before and after transformation is shown in Table\,\ref{tab: astrometry}. We performed a Monte Carlo simulation to determine that the uncertainty of the transformation is dominated by the 0\farcs4 uncertainty on the ATPMN absolute astrometry.

{The \emph{Gaia} EDR3 J2000 proper motion corrected position of \exo\,is 04$^{\rm h}$08$^{\rm m}$07\fhs0986 $-$71$^{\circ}$27\arcmin00\farcs852.}
{The uncorrected position of the radio detection on 2018 August 07 00:04:13.7 is 04$^{\rm h}$08$^{\rm m}$07\fhs1$\pm$0\farcs8 $-$71$^{\circ}$26\arcmin59\farcs3$\pm$0\farcs8. We then applied the absolute astrometry correction where the change in position of the source due only to the absolute astrometry correction was 0\farcs2.
We then applied a proper motion correction to the radio position using the \emph{Gaia} proper motion values to find the J2000 position of the source to be: 04$^{\rm h}$08$^{\rm m}$07\fhs1$\pm$0\farcs9 $-$71$^{\circ}$27\arcmin01\farcs0$\pm$0\farcs5 where we have added the position uncertainty and the 0\farcs4 astrometric uncertainty in quadrature.  The separation between the J2000 corrected 2018 August 07 00:04:13.7 radio position and the J2000 corrected \emph{Gaia} position is 0\farcs2, which is within the uncertainties on the \exo\,radio position. \exo\,is the only source identified by Vizier\footnote{\href{http://vizier.cfa.harvard.edu/viz-bin/VizieR}{http://vizier.cfa.harvard.edu/viz-bin/VizieR}} \citep{vizierpaper} within 5\asec\,of the corrected radio position.}

\section{Optical photometry}
\label{sec: EXO optical photometry}

\exo\,has been observed as part of various optical surveys such as \emph{Gaia}, the All-Sky Automated Survey for Supernovae \citep[ASAS-SN;][]{2018MNRAS.477.3145J}, and the Deep-Near Infrared Survey of the Southern Sky \citep[DENIS;][]{2000A&AS..144..235C}. 

\subsection{Spectral energy distribution}
\label{sec: EXO SED}

We used archival photometry from Vizier \citep{vizierpaper} to construct a spectral energy distribution (SED, the full list of references can be found in Appendix\,\ref{app: SED refs}), shown in Figure\,\ref{fig: EXO SED}. The {\sc bt-settl} \citep[][]{2011ASPC..448...91A,2014IAUS..299..271A} model shown was fit to the SED using the software described in \citet{2012MNRAS.427..343M,2017MNRAS.471..770M} and the \emph{Gaia} DR2 distance of $57.81\pm0.07$\,pc was used to scale the luminosity. The model has temperature of 4000\,K and a surface gravity (log$\left( g \right)$ of 4.5). We can see that there is a UV excess, this is likely due to active regions in the stellar atmosphere \citep[e.g.][]{2011UVexcess}.

\begin{figure*}
\centering
 \includegraphics[width=\textwidth]{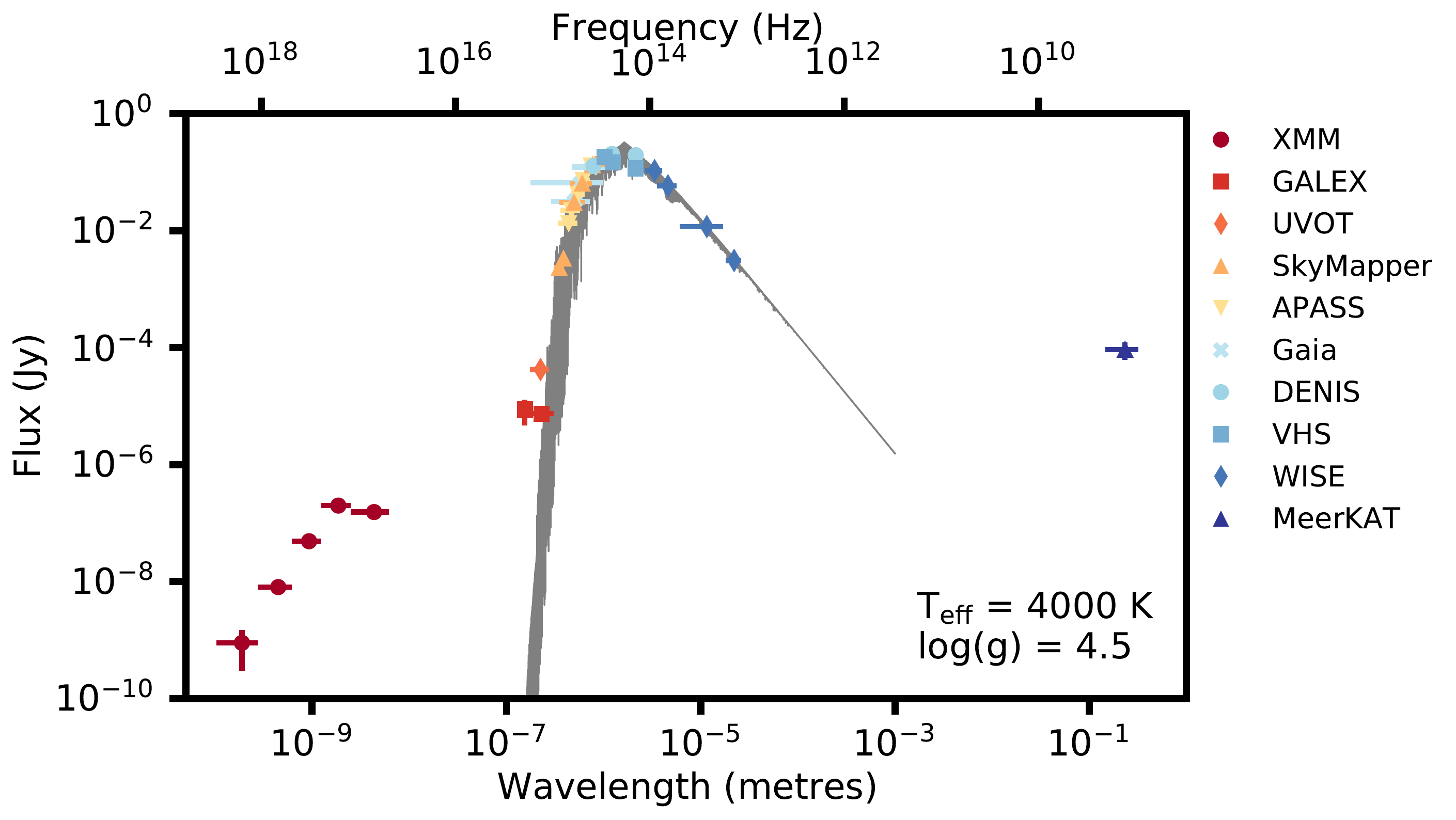}
 \caption{SED of \exo\,using the literature data points (see Appendix\,\ref{app: SED refs}), including the MeerKAT detection and the X-ray detections in Table\,\ref{tab: EXO XMM}. The grey line is the {\sc bt-settl} model fit to the data points, with $T_{\mathrm{eff}}=4000$\,K, $\mathrm{log}_{10}(g)=4.5$, $\mathrm{[Fe/H]}=0$ and $\mathrm{[\alpha/Fe]}=0$.}
 \label{fig: EXO SED}
\end{figure*}

\subsection{Optical variability}
\label{sec: EXO optical var}

\exo\,was classified as a rotational variable star by ASAS-SN \citep{2018MNRAS.477.3145J}, and a variable star, NSV\,15914, by the General Catalogue of Variable Stars \citep{2017ARep...61...80S}.
The source was observed 580 times by the All-sky Automated Survey \citep[ASAS;][]{1997AcA....47..467P} from November 2000 to December 2009, by ASAS-SN \citep[][]{2014ApJ...788...48S,2017PASP..129j4502K} 345 times in $V$-band from 2017 June to 2018 September, and in $g$-band 2180 times from September 2017 to February 2021. We performed a Lomb-Scargle \citep{1976Ap&SS..39..447L, 1982ApJ...263..835S,2018MNRAS.478..218P} analysis on the ASAS, ASAS-SN $V$-band, and ASAS-SN $g$-band observations. We did not find a period for the original ASAS observations due to the sparsity of the data points. 
Both the ASAS-SN $V$- and $g$-band observations are affected by the approximately daily observing cadence; however, we find a period of $5.13\pm0.03$\,days for the ASAS-SN $V$-band observations, and a period of $5.20\pm0.02$\,days for the ASAS-SN $g$-band observations. These are consistent with the $5.2\pm0.20$\,day period found by \citet{1996_EXO_firstPeriod}. The median $V$-band magnitude is 12.4\,mag from both ASAS and ASAS-SN, and the median ASAS-SN $g$-band magnitude is 13.1\,mag.

\subsection{\emph{TESS} optical photometry}
\label{sec: EXO TESS obs}

\exo\,has been observed in 13 Transiting Exoplanet Survey Satellite \citep[\emph{TESS};][]{2015JATIS...1a4003R} sectors. \emph{TESS} first observed the source in July 2018 and the last observations were taken in July 2019. \citet{2020_EXO_TESS_Flares} found 11 flares from \exo\,in an automated search for stellar flares from M-dwarfs in the first 2 \emph{TESS} sectors. We can see in Figure\,\ref{fig: EXO MKT TESS lc} that there is no optical flare during the MeerKAT observations. From the \emph{TESS} observations, \citet{2020_EXO_TESS_Flares} also found a rotational period of 4.92\,days (no uncertainty is given), which is broadly consistent with the $5.2\pm0.20$\,day period. 

\begin{figure}
\centering
 \includegraphics[width=\columnwidth]{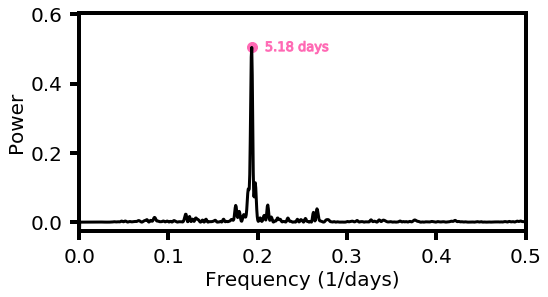}
 \caption{{\emph{TESS} Lomb-Scargle periodogram generated using the clean, de-trended data. The 5.18\,day period is indicated.}}
 \label{fig: EXO TESS periodogram}
\end{figure}

\begin{figure*}
\centering
 \includegraphics[width=0.95\textwidth]{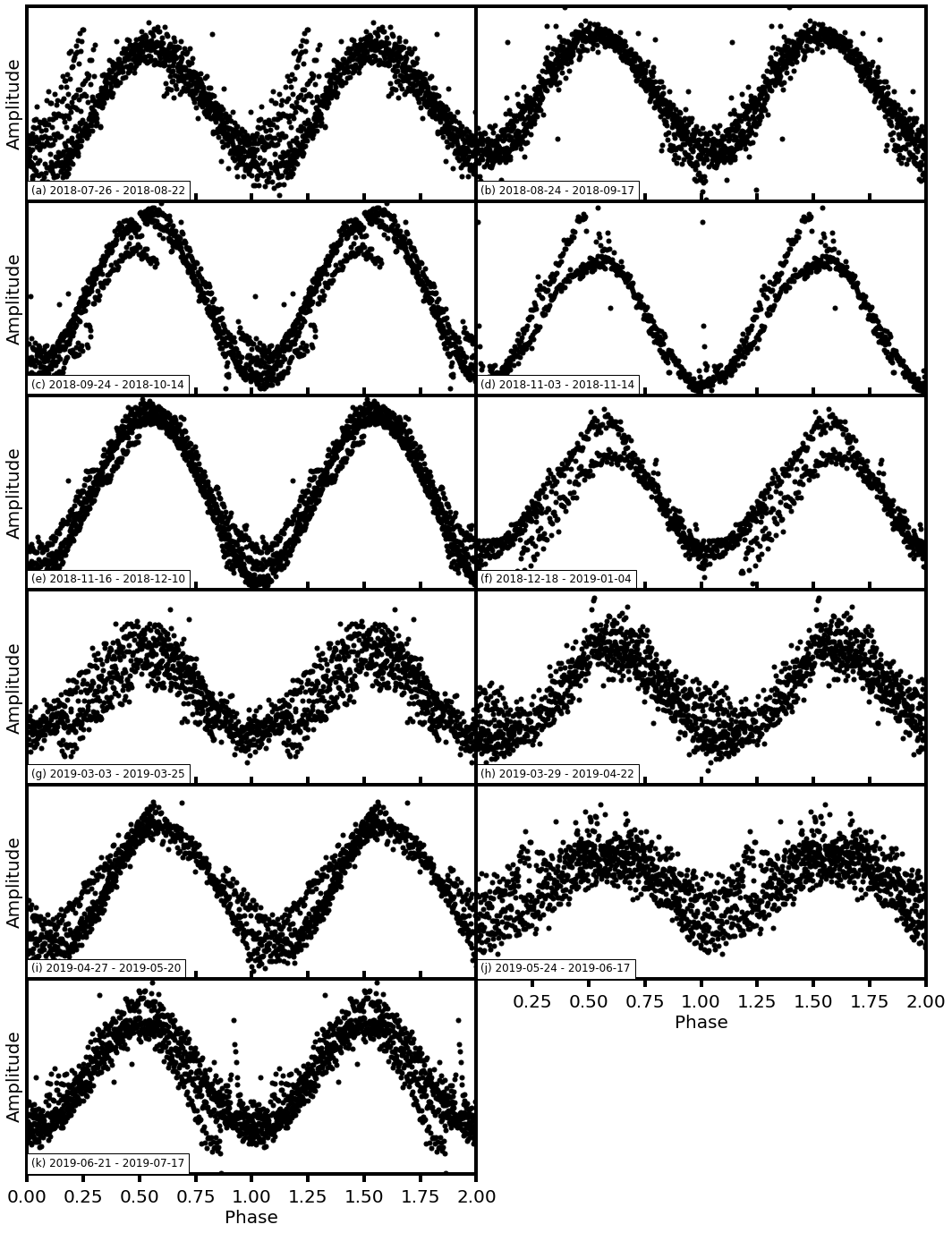}
 \caption{Eleven \emph{TESS} sectors folded to a period of 5.18$\pm$0.04\,days, 2 sectors were not included due to data issues. Note that the amplitude is arbitrary and we have not verified the short, bright bursts as real flares.}
 \label{fig: EXO TESS all folded}
\end{figure*}

The \emph{TESS} Input Catalogue (TIC) Identifier for \exo\,is 25132999. Upon inspection of the \emph{TESS} images, there is a nearby source with coordinates $04^{\mathrm{h}}07^{\mathrm{m}}56\fhs190$ $-71^{\circ}$26\arcmin36\farcs851 and TIC 25132995 that is partially blended with \exo. We extracted \emph{TESS} light curves for this source as well as \exo\, and another nearby source, TIC 677356964 with coordinates $04^{\mathrm{h}}08^{\mathrm{m}}44\fhs444$ $-71^{\circ}$28\arcmin49\farcs696. We extracted the light curve for TIC 25132995 to determine whether the periodicity found for \exo\,has been affected or induced by this source. We extracted the light curve for TIC 677356964 to find and eliminate systematic problems with the light curves. We removed any part of the light curve where at least two of the three sources have correlated systematic issues, and removed two sectors where the data were strongly affected by systematics. We then divided each sector into parts where there was a gap in observations of two days or more, and for each part we removed any linear trend in the data.  We performed a Lomb-Scargle analysis on all of the clean, de-trended data and found a period of 5.18$\pm$0.04\,days. {The Lomb-Scargle periodogram is shown in Figure\,\ref{fig: EXO TESS periodogram}} When we fold the data to 5.18\,days, shown in Figure\,\ref{fig: EXO TESS all folded}, we find that the shape of the curve and phase of the peak is consistent across the year of \emph{TESS} observations. Due to the challenge of calibrating the absolute flux of \emph{TESS} observations, the amplitude of the observations shown here are arbitrary and not scaled relative to each sector. We have not verified the nature of the bright, short bursts.

\section{Optical spectroscopy}
\label{sec: EXO optical spectroscopy}

\begin{landscape}
\begin{figure}
\centering
 \includegraphics[width=\linewidth]{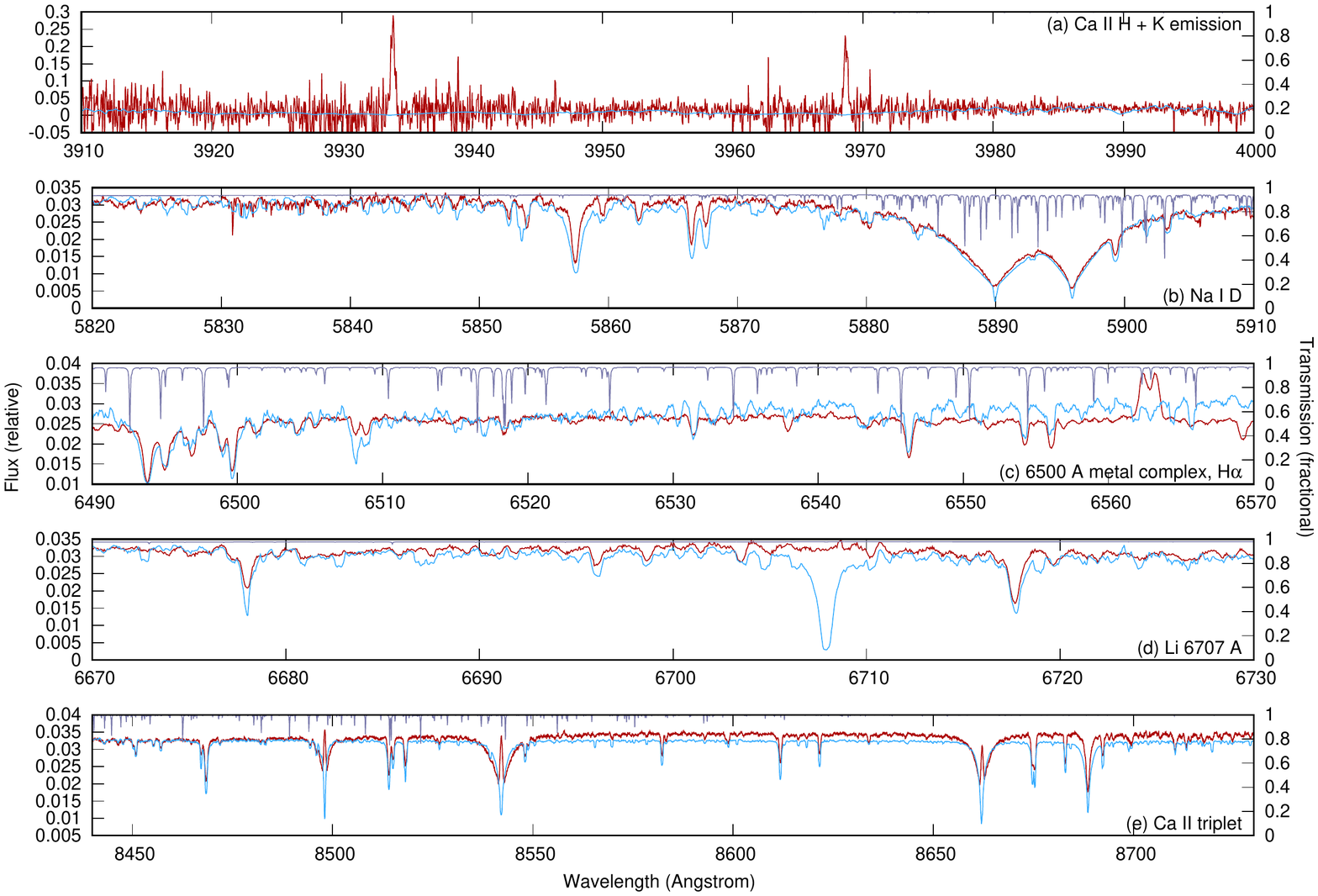}
 \caption[SALT HRS spectrum of \exo.]{SALT HRS spectrum of \exo. The red lines are the spectrum of \exo, the cyan lines are the {same model spectrum used in Figure \ref{fig: EXO SED}, and the purple lines show the atmospheric transmission. The spectrum has been corrected for a radial velocity of 16\,km\,s$^{-1}$. Panel (a) shows the Ca\,{\sc ii} H \& K lines, panel (b) shows the Na\,{\sc i}\,D lines, panel (c) shows the 6500\,\AA\,metal line complex and H$\alpha$ (double-peaked emission line), panel (d) shows the Li 6707\,\AA\ line, and panel (e) shows the Ca\,{\sc ii} triplet.
 }}
 \label{fig: EXO SALT spec}
\end{figure}
\end{landscape}

We observed \exo\,with the Southern African Large Telescope \citep[SALT;][]{Buckley2006} High Resolution Spectrograph \citep[HRS;][]{Bramall2012,Crause2014} on 2021 February {17.7835} {in clear conditions and $\sim$1\arcsec seeing}. SALT is located at the South African Astronomical Observatory (SAAO) site in Sutherland, South Africa, and HRS is a fibre-fed, dual-beam, white-pupil, vacuum-stabilised, high-resolution (R = 13,000$-$74,000, depending on mode and wavelength) spectrograph. The HRS spectrum was taken in low-resolution mode with an integration time of 1200\,seconds. Blue (3800$-$5550\AA) and red (5450$-$8900\AA) spectra were obtained {at average resolutions of 0.3\AA\/ and 0.5\AA, respectively}, and were calibrated using the weekly set of HRS calibrations, including ThAr arc spectra and QTH lamp flat-fields.
{Two identical fibres sampled the object and a nearby (20 arcsec) sky region, producing interleaving {\`e}chelle orders, which were reduced using the SALT HRS MIDAS pipeline \citep{Kniazev2016} to produce a sky subtracted spectrum, a sample of which is shown in Figure\,\ref{fig: EXO SALT spec}.}

The spectrum confirms that the source is a chromospherically active M-dwarf with a temperature of $4000\pm$\,$\sim$\,200\,K. The chromospheric activity is identified using the H$\alpha$ line, which is in emission with a filled core (see Figure\,\ref{fig: EXO SALT spec}(c)) and the calcium H \& K lines and near-infrared calcium triplet, which are in absorption with strong core emission (see Figure\,\ref{fig: EXO SALT spec}(a) and (e)). In Figure\,\ref{fig: EXO SALT spec}(d) we can see that there is no lithium absorption and we do not see any lines suggesting a binary companion.

\section{X-ray and UV observations}
\label{sec: EXO X-ray}
\label{sec: EXO UV}

As well as the original \textit{EXOSAT} observations, \exo\,has been observed by \textit{XMM-Newton} and \textit{ROSAT}. The source was faintly detected in the \textit{ROSAT} All-Sky-Survey \citep{1999A&A...349..389V} and was detected once serendipitously in the \textit{XMM-Newton} slew survey on 2011 September 22 \citep{2018yCat.9053....0X}. The mean flux and luminosity in each \textit{XMM-Newton} band is shown in Table\,\ref{tab: EXO XMM}.

\begin{table}
    \centering
    \begin{tabular}{l|r|r}
    Band (keV) & Flux ($\mathrm{mW\,m^{-2}}$) & Luminosity ($\mathrm{erg\,s^{-1}}$) \\
    \hline 
    0.2-0.5 & $(1.1170\pm 0.0371)\times10^{-13}$    & $(4.5\pm0.1)\times10^{28}$ \\
    0.5-1.0 & $(2.4021\pm0.0547)\times10^{-13}$     & $(9.6\pm0.2)\times10^{28}$ \\
    1.0-2.0 & $(1.1804\pm0.0417)\times10^{-13}$     & $(4.7\pm0.2)\times10^{28}$ \\
    2.0-4.5 & $(0.4850\pm0.0493)\times10^{-13}$      & $(1.9\pm0.2)\times10^{28}$ \\
    4.5-12  & $(0.1620\pm0.108)\times10^{-13}$       & $(0.6\pm0.4)\times10^{28}$ \\
    0.2-12  & $(5.4914\pm0.152)\times10^{-13}$      & $(22.0\pm0.6)\times10^{28}$ \\
    0.5-4.5 & $(5.2819\pm0.102)\times10^{-13}$      & $(21.1\pm0.4)\times10^{28}$ \\
    \hline
    \end{tabular}
    \caption[\textit{XMM-Newton} bands and fluxes for \exo.]{\textit{XMM-Newton} bands and fluxes for \exo.}
    \label{tab: EXO XMM}
\end{table}

VW\,Hyi was monitored by \textit{Swift} as part of the ThunderKAT observations; however, \exo\,did not fall into the FoV of the X-ray Telescope (\textit{Swift} XRT). \exo\,did fall within the {Ultraviolet/Optical Telescope} \citep[UVOT;][]{2000HEAD....5.3411M} FoV during these observations, on 2018 July 31 while observing with the UM2 filter (centred on 2246\,\AA) {and it} was detected with an AB\,magnitude of 19.84$\pm$0.21. \exo\,has also been detected in Galaxy Evolution Explorer Data Release 5 \citep[GALEX-DR5;][]{2011Ap&SS.335..161B}, with FUV (1344-1786\,\AA) and NUV (1771-2831\,\AA) AB\,magnitudes of 21.544 and 20.176 respectively.

\section{Discussion}
\label{sec: EXO discussion}

\begin{figure}
\centering
 \includegraphics[width=\columnwidth]{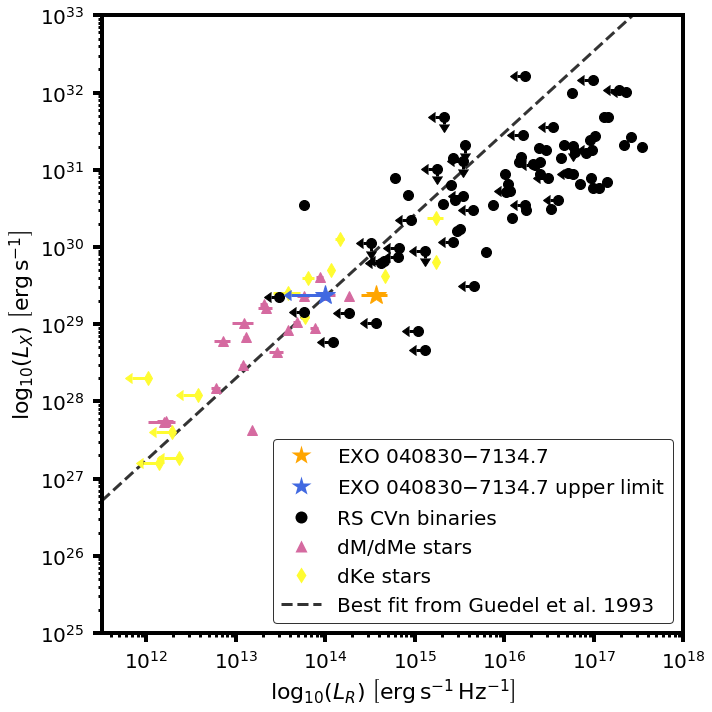}
 \caption[\gudel\,plot including \exo.]{Radio specific quiescent luminosities and X-ray quiescent luminosities of stars (\gudel\,plot) showing the sources from \citet{1993ApJ...405L..63G} and \exo. The blue \exo\,upper limit is the radio upper limit from the first epoch, which assumes that the detection in the second epoch is a flare. The orange \exo\,point is the flux density in the second epoch, which assumes the detection is the quiescent radio emission from the source. The \gudel--Benz relation is shown using a dashed-line.}
 \label{fig: EXO Benz-Guedel}
\end{figure}

\exo\,is the first radio M-dwarf detected with MeerKAT. We have placed the source on the \gudel\,plot (Figure\,\ref{fig: EXO Benz-Guedel}, the \gudel--Benz relation, shown as a dashed line) using the \textit{EXOSAT} quiescent X-ray flux, adjusted for the new 57.81\,pc distance. {We can see that the more recent \textit{XMM-Newton} X-ray luminosities (Table\,\ref{tab: EXO XMM}) are consistent with the \textit{EXOSAT} luminosity.}
The blue upper limit assumes that the detection in the second MeerKAT epoch is a flare, and uses the non-detection upper limit on the radio flux density from the full-time-integration image of the first epoch. The orange marker assumes that the detection in the second epoch is the quiescent radio flux density of the source.
These are only estimates of the position of the source on the plot, as the quiescent radio and X-ray emission should ideally be measured simultaneously.  We can see that in both cases \exo\,fits well with the relation and with the M-dwarfs in the plot.

{We can determine limits on the brightness temperature, $T_b$, of the radio detection of \exo\,using the formula:
\begin{eqnarray}
T_b = \frac{c^2}{\kappa \nu^2} \cdot F_{\nu} \cdot \frac{d^2}{A^2}
\end{eqnarray}
\citep{1985ARA&A..23..169D} where $c$ is the speed of light, $\kappa$ is the Boltzmann constant, $\nu$ is the observing frequency, $F_{\nu}$ is the flux density at frequency $\nu$, $d$ is the distance to the source, and $A$ is the projected area of the emitting region where $A=\pi R_{*}^{2}$. If we assume a typical M-dwarf stellar radius of $0.6\,M_{\odot}$ and that the emitting radio region is twice the radius of the star, then $A=1.5\times10^{22}\,\mathrm{cm}$. Using the MeerKAT 1284\,MHz flux density of the \exo\,when it was detected on 2018 August 07 ($0.09\pm0.03\,\mathrm{mJy}$) we find the corresponding brightness temperature to be $T_{b}\sim2\times10^9\,\mathrm{K}$, which is consistent with an incoherent emission mechanism. This is likely a lower limit on the brightness temperature, as we are assuming a large emitting region. If we assume instead a plasma loop with radius $\sim7\times10^8\,\mathrm{cm}$ \citep{2001A&A...374.1072S}, the brightness temperature could reach $T_{b}\sim2\times10^{13}\,\mathrm{K}$, which would imply a coherent emission mechanism. Polarisation information and a radio spectrum would assist in determining the type radio emission; however, the polarisation properties of MeerKAT images away from the phase centre are still being categorised and data processing methods developed. 
As such, polarisation studies of this source will be investigated as part of future work. 
The source is insufficiently bright to measure the radio spectrum or to determine any meaningful constraints. The estimated brightness temperature of \exo\, (assuming a large emission region), the position on the \gudel--Benz plot (Figure\,\ref{fig: EXO Benz-Guedel}) and the chromospheric activity indicated by the H$\alpha$, calcium H \& K lines, and near-infrared calcium triplet suggest that the radio emission is incoherent gyrosynchrotron emission. 
}

The SED, optical spectrum, and photometric variability support the conclusion that \exo\,is a magnetically active, rotating, single M-dwarf and that the radio emission that we observed with MeerKAT is either an incoherent flare or quiescent emission.
We used \emph{TESS} photometry to reveal that the rotational period of \exo\,is 5.18$\pm$0.04\,days, and to show that there is no optical flare at the time of the radio observations. Simultaneous optical and radio monitoring of Proxima Centauri by \citet{2020ApJ...905...23Z} detected an optical flare a few hours before the detection of a radio flare and they found that the optical and radio emission were from a Type IV burst linked to a coronal mass ejection. However; optical and radio bursts are not necessarily observed together and can be produced by different flaring mechanisms. 

{\exo\,is only the first MeerKAT detection of a radio star, but we can use recent searches for stellar radio emission to determine whether this detection is consistent with the expected rates. The FWHM of MeerKAT is $\sim1\,\mathrm{deg^{-2}}$, we therefore have a surface density of radio detected M-dwarfs of $1\,\mathrm{deg^{-2}}$ at 1284\,MHz with a sensitivity of $0.03\,\mathrm{mJy\,beam^{-1}}$. \citet{2021MNRAS.502.5438P} searched ASKAP Stokes \textit{I} and \textit{V} images at 887.5\,MHz and found radio emission from 33 stars, 23 of which had no previous radio detections. They determine that the surface density of radio loud cool dwarfs stars at 887.5\,MHz is $5.27^{+3.06}_{-2.15}\times10^{-4}\,\mathrm{deg^{-2}}$ with a sensitivity of $0.25\,\mathrm{mJy\,beam^{-1}}$. Using the same scaling relation as \citet{2021MNRAS.502.5438P} ($N(>S)\propto S^{-3/2}$ where $N$ is the number of stars and $S$ is the flux density) the ASKAP result scales to $0.01\mathrm{deg^{-2}}$ for MeerKAT. Using the Faint Images of the Radio Sky at Twenty Centimetres \citep[FIRST;][]{FIRST1995} survey \citet{1999AJ....117.1568H} searched $\sim5000\,\mathrm{deg^{2}}$ with a sensitivity of $0.07\,\mathrm{mJy\,beam^{-1}}$ at 1400\,MHz and found 26 stellar radio sources. This corresponds to a surface density of $5.2\times10^{-3}\,\mathrm{deg^{-2}}$, which scales to $0.6\,\mathrm{deg^{-2}}$ using the sensitivity of the MeerKAT observations of \exo.
As we search more MeerKAT observations for stellar sources our statistics will become more robust, but using this one source our rate is more consistent with the FIRST rate than the ASKAP rate. This could be because cool dwarfs are more active at $\sim1400\,\mathrm{MHz}$ than at $\sim900\,\mathrm{MHz}$, or it could be due to the different search methods. We searched for radio emission at the position of a known active star and \citet{1999AJ....117.1568H} matched the position of radio sources to the known positions of stars. Whereas \citet{2021MNRAS.502.5438P} searched specifically for circularly polarised searches. This may mean that they are missing radio stars with circular polarisation fractions below their threshold. The MeerKAT Galactic Plane Survey will be particularly interesting for investigating the distribution and surface density of stellar radio sources \citep[][]{2016mks..confE..15T}.}

\section{Conclusions}
\label{sec: EXO conclusions}

We observe radio emission from the known X-ray flaring star \exo\,during MeerKAT observations of the CV VW Hyi.  We use the SED and new SALT spectroscopy of the source to confirm that it is an M-dwarf, with a temperature of $4000\pm200$\,K and spectral emission lines indicating chromospheric activity. We use \emph{TESS} optical photometry to determine that the rotational period of the star is 5.18$\pm$0.04\,days.
This is the first radio emission detected from the star, and the first MeerKAT detection of an M-dwarf flare star.
{The radio brightness temperature, the X-ray and radio luminosity, and the chromospheric activity of the star indicate that the radio emission is incoherent gyrosynchrotron radiation.}
This detection shows the importance of commensal searches for variable and transient radio sources, and their potential for finding more radio stars.

\section*{Acknowledgements}

LND and BWS acknowledge support from the European Research Council (ERC) under the European Union's Horizon 2020 research and innovation programme (grant agreement No 694745). 
DRAW was supported by the Oxford Centre for Astrophysical Surveys, which is funded through generous support from the Hintze Family Charitable Foundation.
DAHB acknowledges research support from the South African National Research Foundation. 
We acknowledge use of the Inter-University Institute for Data Intensive Astronomy (IDIA) data intensive research cloud for data processing. IDIA is a South African university partnership involving the University of Cape Town, the University of Pretoria and the University of the Western Cape.
The MeerKAT telescope is operated by the South African Radio Astronomy Observatory (SARAO), which is a facility of the National Research Foundation, an agency of the Department of Science and Technology. We would like to thank the operators, SARAO staff and ThunderKAT Large Survey Project team. The Parkes radio telescope is part of the Australia Telescope National Facility which is funded by the Commonwealth of Australia for operation as a National Facility managed by CSIRO. Some of these observations were obtained with the Southern African Large Telescope under the Large Science Programme on transients, 2018-2-LSP-001 (PI: DAHB).
This work has made use of data from the European Space Agency (ESA) mission
\emph{Gaia}.\footnote{\href{https://www.cosmos.esa.int/gaia}{https://www.cosmos.esa.int/gaia}, processed by the \emph{Gaia}
Data Processing and Analysis Consortium (DPAC,
\href{https://www.cosmos.esa.int/web/gaia/dpac/consortium}{https://www.cosmos.esa.int/web/gaia/dpac/consortium})} Funding for the DPAC
has been provided by national institutions, in particular the institutions
participating in the \emph{Gaia} Multilateral Agreement.
This research made use of Astropy,\footnote{http://www.astropy.org} a community-developed core Python package for Astronomy \citep{astropy:2013, astropy:2018}.
This research made use of APLpy, an open-source plotting package for Python \citep{2012ascl.soft08017R}.

\section*{Data availability}

The data underlying this article are available in Zenodo at \href{https://doi.org/10.5281/zenodo.5084298}{https://doi.org/10.5281/zenodo.5084298}.
The code underlying this article is available on GitHub and Zenodo:
\href{https://github.com/IanHeywood/oxkat}{https://github.com/IanHeywood/oxkat}, \href{https://doi.org/10.5281/zenodo.4515114}{https://doi.org/10.5281/zenodo.4515114}, \href{https://doi.org/10.5281/zenodo.4456303}{https://doi.org/10.5281/zenodo.4456303}
and
\href{https://doi.org/10.5281/zenodo.4921715}{https://doi.org/10.5281/zenodo.4921715}.




\bibliographystyle{mnras}
\bibliography{LongTermVariables,EXO} 



\appendix

\section{Supplementary material}
\label{app: SED refs}

The sources for the optical SED shown in Figure\,\ref{fig: EXO SED} are:
\begin{itemize}
    \item \emph{Gaia} Data Release 2. Summary of the contents and survey properties \citep{2018A&A...616A...1G}
    \item VizieR Online Data Catalog: \emph{Gaia} Early Data Release 3 \citep[EDR3;][]{2020yCat.1350....0G}
    \item VizieR Online Data Catalog: Wide-field Infrared Survey Explorer (WISE) All-Sky Data Release \citep{2012yCat.2311....0C}
    \item Galaxy Evolution Explorer (GALEX) catalogs of UV sources: statistical properties and sample science applications: hot white dwarfs in the Milky Way \citep{2011Ap&SS.335..161B}
    \item American Association of Variable Star Observers (AAVSO) Photometric All-Sky Survey (APASS) - The Latest Data Release \citep{2015AAS...22533616H}
    \item SkyMapper Southern Survey: First Data Release \citep{2018PASA...35...10W}
    \item The Deep-Near Infrared Survey of the Southern Sky \citep[DENIS;][]{2000A&AS..144..235C}
    \item The Visible and Infrared Survey Telescope for Astronomy (VISTA) Hemisphere Survey \citep[VHS;][]{2013Msngr.154...35M}
\end{itemize}


\bsp	
\label{lastpage}
\end{document}